\documentclass{aastex63}
\usepackage{graphicx}
\usepackage{amsmath}
\usepackage{amssymb}
\usepackage{chemformula}
\usepackage{natbib}
\submitjournal{PSJ}
\shorttitle{A Microphysical Cloudy Clear Climate Model}
\shortauthors{Windsor et al.}
\begin{document}

\title{A Radiative-Convective Model for Terrestrial Planets with Self-Consistent Patchy Clouds}

\correspondingauthor{James D. Windsor}
\email{jdw472@nau.edu}

%
\author[0000-0001-8522-3788]{James D. Windsor}
\affiliation{Department of Astronomy and Planetary Science, Northern Arizona University, Flagstaff, AZ 86011, USA}
\affiliation{Habitability, Atmospheres, and Biosignatures Laboratory, University of Arizona, Tucson, AZ 85721, USA}
\affiliation{NASA Nexus for Exoplanet System Science Virtual Planetary Laboratory, University of Washington, Box 351580, Seattle, WA 98195, USA}

\author[0000-0002-3196-414X]{Tyler D. Robinson}
\affiliation{Lunar \& Planetary Laboratory, University of Arizona, Tucson, AZ 85721 USA}
\affiliation{Department of Astronomy and Planetary Science, Northern Arizona University, Flagstaff, AZ 86011, USA}
\affiliation{Habitability, Atmospheres, and Biosignatures Laboratory, University of Arizona, Tucson, AZ 85721, USA}
\affiliation{NASA Nexus for Exoplanet System Science Virtual Planetary Laboratory, University of Washington, Box 351580, Seattle, WA 98195, USA}

\author[0000-0002-5893-2471]{Ravi kumar Kopparapu}
\affiliation{NASA Goddard Space Flight Center 8800 Greenbelt Road Greenbelt, MD 20771, USA}

\author[0000-0003-4580-3790]{David E. Trilling}
\affiliation{Department of Astronomy and Planetary Science, Northern Arizona University, Flagstaff, AZ 86011, USA}

\author[0000-0003-4450-0368]{Joe LLama}
\affiliation{Lowell Observatory, 1400 W. Mars Hill Rd. Flagstaff. Arizona. 86001. USA.}

\author[0000-0003-3099-1506]{Amber Young}
\affiliation{Department of Astronomy and Planetary Science, Northern Arizona University, Flagstaff, AZ 86011, USA}
\affiliation{Habitability, Atmospheres, and Biosignatures Laboratory, University of Arizona, Tucson, AZ 85721, USA}
\affiliation{NASA Nexus for Exoplanet System Science Virtual Planetary Laboratory, University of Washington, Box 351580, Seattle, WA 98195, USA}

%

%
\begin{abstract}

Clouds are ubiquitous\,---\,they arise for every solar system planet that possesses an atmosphere and have also been suggested as a leading mechanism for obscuring spectral features in exoplanet observations. As exoplanet observations continue to improve, there is a need for efficient and general planetary climate models that appropriately handle the possible cloudy atmospheric environments that arise on these worlds. We generate a new 1D radiative-convective terrestrial planet climate model that self-consistently handles patchy clouds through a parameterized microphysical treatment of condensation and sedimentation processes. Our model is general enough to recreate Earth's atmospheric radiative environment without over-parameterization, while also maintaining a simple implementation that is applicable to a wide range of atmospheric compositions and physical planetary properties. We first validate this new 1D patchy cloud radiative-convective climate model by comparing it to Earth thermal structure data and to existing climate and radiative transfer tools. We produce partially-clouded Earth-like climates with cloud structures that are representative of deep tropospheric convection and are adequate 1D representations of clouds within rocky planet atmospheres. After validation against Earth, we then use our partially clouded climate model and explore the potential climates of super-Earth exoplanets with secondary nitrogen-dominated atmospheres which we assume are abiotic. We also couple the partially clouded climate model to a full-physics, line-by-line radiative transfer model and generate high-resolution spectra of simulated climates. These self-consistent climate-to-spectral models bridge the gap between climate modeling efforts and observational studies of rocky worlds.  

\end{abstract}
%

%
%

%

%

\section{Introduction}

The field of exoplanet astronomy continues to expand rapidly since the first discovery of a world orbiting a Sun-like star \citep{mayorandqueloz1995}. Currently there are more than 5,000 confirmed exoplanets\footnote{https://exoplanetarchive.ipac.caltech.edu} and planet occurrence rate studies \citep{howardetal2012,fressinetal2013,Beanetal2021} suggest that so-called super-Earth and sub-Neptune worlds are the most common planet type in the galaxy. For these worlds, exoplanet mass-radius data reveal that worlds smaller than roughly 1.6 Earth radii are likely to be rocky \citep{rogers2015}. 
Occurrence rates for small rocky worlds orbiting within the Habitable Zone \citep{kastingetal1993,kopparapuetal2013} of their host star remain uncertain due to survey completeness issues, but synthesis \footnote{https://exoplanets.nasa.gov/exep/exopag/sag/\#sag13} studies indicate that such ``exo-Earth candidates'' are not uncommon \citep{starketal2014,dressingandcharbonneau2015,starketal2015,Kopparapuetal2018}. The eventual characterization of Earth-like exoplanets\,---\,for both transiting and directly-imaged worlds\,---\,remains a primary motivator of exoplanet astronomy\footnote{https://www.nationalacademies.org/our-work/decadal-survey-on-astronomy-and-astrophysics-2020-astro2020}. Fully characterizing an exoplanet atmosphere requires a deep understanding of underlying physical and chemical properties and the ability to connect these properties to observed spectral signatures. Thus, the successful study of potentially habitable worlds will rely heavily on the development of a hierarchical suite of exoplanet climate models that span appropriate ranges of complexity and computational expense \citep{fouchezetal2021}.

One leading problem in the field of exoplanet science is the spectral influence of clouds in exoplanet observations \citep{hellingetal2019}. A main driver of current planetary atmosphere modeling efforts is connecting these cloud-induced spectral signatures to underlying atmospheric states. Clouds and aerosols have a strong influence on the overall energy budget of a planetary atmosphere and can have profound effects on the resulting planetary climate \citep{pierrehumbert2011}. Likewise, clouds also dramatically influence the spectral signatures of exoplanetary atmospheres where, in a number of observations \citep[e.g.,][]{pontetal2008,kreidbergetal2014a,knutsonetal2014,singetal2015,diamondloweetal2018}, a muting (or loss) of spectral features is due to the likely presence of high-altitude, vertically-extended clouds. Finally, understanding how clouds impact spectral observations and exoplanet atmospheric structure is of critical importance to the full characterization of nearly all worlds \citep{hellingetal2019}.

A multitude of aerosol tools have been used to explore cloud physics in exoplanet atmospheres, ranging in complexity from highly-parameterized cloud models \citep[e.g.,][]{rossow1978,barstowetal2017,tsiarasetal2018,fisherandheng2018,pinhasetal2019,Barstow2020} to sophisticated microphysical tools like the Community Aerosol and Radiation Model for Atmosphere \citep[\texttt{CARMA};][]{toonetal1988,murphyetal1993,Zhao1995,gaoetal2014a,Powelletal2019,Rooney2022}. In practice, model complexity is selected through balancing computational expense against the potential limitations of underlying model parameterizations.
Simple cloud models generally adopt a small number of free parameters (e.g., cloud top pressure, total cloud optical depth) and are well-suited to interpreting noisy observations \citep{Barstow2020}. Intermediate-complexity cloud models can include some additional physical processes that control aerosol particle size distributions and vertical extent \citep[e.g., sedimentation physics;][]{ackermanandmarley2001,morleyetal2012}. The parameterization of some processes, such as prescribing the shape of the cloud droplet size distribution, then enables incorporation of intermediate-complexity cloud tools into both forward climate models \citep{marleyetal2010,ZSOM2012,morleyetal2012} and retrieval models \citep{maiandline2019}.

The most complex cloud models contain detailed treatments of cloud microphysical processes and explicitly resolve condensate size distributions \citep{gaoetal2021}. Such aerosol models treat microphysical processes kinetically, producing size- and altitude-resolved particle distributions by considering processes such as diffusive mixing, sedimentation, coagulation, condensation, and evaporation \citep{gaoetal2021}. In general, these more complex models are used to explore how different condensate species interact with each other and how condensate size distributions form and evolve through exchanges with the surrounding atmospheric medium \citep{toonetal1988,murphyetal1993,Zhao1995,gaoetal2014a,Powelletal2019,Rooney2022}. Applications of full-physics 1D microphysical models to exoplanet atmospheres \citep{gaoetal2014a,Powelletal2019,gaoetal2021} have been instrumental in demonstrating the strengths and limitations of lower-complexity aerosol models.

The incorporation of cloud tools across the complexity spectrum into planetary climate models has enabled novel studies of planetary energy balance and transport. For example, 3D General Circulation Models (GCMs) with adopted cloud parameterizations have shown that extensive upwelling motions on the day side of a rotationally-locked Earth-like exoplanet may lead to wide-scale cloudiness over the illuminated portion of the planet \citep{pierrehumbert2011,wayetal2017,checlairetal2019}. One-dimensional (vertical) radiative-convective models have also incorporated cloud treatments\,---\,especially for brown dwarfs and gas giant exoplanets \citep{marleyetal2010,fortneyetal2011,morleyetal2012}\,---\,and such tools are an important component in the hierarchy of planetary climate models \citep{fouchezetal2021} owing to their versatility and computational speed. To date, applications of cloud models inside 1D terrestrial planet climate tools has been limited. A limited number of patchy-cloud 1D radiative-convective models have been explored for Earth-like conditions \citep{Kitzmann2010,Kitzmann2011,ZSOM2012,fouchezetal2018}, but extending these models to less Earth-like scenarios risks issues with tuned parameterization  \citep{selsisetal2007,marleyetal2010,ZSOM2012,Kitzmann2010}.


The work presented below explores the creation of a general tool for simulating 1D radiative-convective climate states for rocky exoplanets with patchy clouds. Importantly, this tool builds upon a successful 1D cloud-free rocky exoplanet climate model while incorporating a treatment of patchy clouds pioneered in the brown dwarf and giant exoplanet literature.  In Section~\ref{sec:methods}, we present the tools that underlay our new partially-clouded 1D radiative-convective climate model, including a detailed description of the cloud microphysics, sub-grid cloudy and clear radiative transfer, and an overview of the basic climate model. In Section~\ref{sec:Model Validation and Sensitivity}, we validate the newly developed climate model following similar methods to \citet{ZSOM2012}, in which we hone in the most efficient number of model layers, test the effect of water vapor abundance, explore the effect of user prescribed fractional cloudiness and generate an overall baseline Earth-like climate model. In Section~\ref{sec:results} we explore a new method of rapid model exploration via a cloudy ``inverse'' climate model, and select four forward models to investigate as equivalent cloudy modern Earth climates. Section~\ref{sec:discussion} contains a discussion of our findings, including material on cloud radiative forcing, cloudy greenhouse effects, the habitability of super-Earth worlds and a snapshot of future work. Section~\ref{sec:conclusion} presents our conclusions and summarizes key takeaways from this study.

\section{Methods}
\label{sec:methods}

In the most general terms, a climate model is a numerical approximation of a climate system that is based on physical, chemical, and sometimes biological principles. Most of the relations that could be used to predict influences from these basic principles are either overly complex or not fully understood, implying that climate solutions are often numerical and approximate. Most commonly, 3D GCMs \citep{phillips1956,turbetetal2021} or 1D (vertical) climate models \citep{manabeetal1961,manabe&strickler1964,kasting&ackerman1986,kasting1988,kastingetal1993,pierrehumbert1995,pavlovetal2000,seguraetal2007,selsisetal2007,kopparapuetal2013} are adopted to generate numerical climate solutions. In 3D climate tools, the need to model material flows often necessitates heavy parameterization of other processes, such as radiative transfer and cloud physics \citep{wayetal2017}. By comparison, 1D models omit physical treatments of 3D material flows to achieve fast runtimes while often including more-detailed models of radiative transfer and, potentially, cloud physics \citep{gaoetal2021}. Notably, 1D climate models are often used as an initial tool in planetary characterization \citep{seguraetal2005,wordsworthetal2010b,vonparisetal2010,kalteneggeretal2011,rugheimer20etal12,lincowskietal2018}, planetary habitability studies \citep{kastingetal1993,kopparapuetal2013}, and also serve as the basis for many exoplanet spectral models \citep{seguraetal2005,giallucaetal2021}. 

A primary characteristic of climate is the thermal structure of the atmosphere resulting from energy transport interactions through conduction, radiation, and advection. 
In general, heat transport through conductive processes in planetary atmospheres is negligible, leaving most heat transport to take place through radiative and advective mechanisms. When considering 1D models (as done hereafter), with only the vertical dimension represented, advection is limited to vertical convection or mixing (or parameterizations of these). Here, regions in an atmosphere can be characterized as either convective or radiative through comparing the local temperature gradient to an ``adiabatic'' gradient. It is common to define the local gradient ($\nabla$) in terms of the logarithmic rate of change of temperature with pressure,   
\begin{equation*}
    \nabla = \frac{\partial \ln T}{\partial \ln p} \ ,
\end{equation*}
where $T$ is temperature and $p$ is pressure. In atmospheric regions where the local temperature gradient exceeds that of the dry adiabatic gradient, the medium is unstable to convection \citep[absent any stabilizing compositional gradients;][]{tremblinetal2019}. Conversely, where the temperature gradient is smaller than the adiabatic gradient, the medium is stable against vertical convection and energy transport proceeds solely through radiative mechanisms (again, absent any destabilizing compositional gradients). In radiative-convective climate models, unstable layers are either assumed to relax to a prescribed adiabatic structure, where ``convective adjustment'' \citep{manabe&strickler1964,manabe&wetherald1967} or mixing-length theory \citep{vitense1953,gierasch&goody1968} can be used to parameterize the vertical mixing of heat that would occur due to convective motions. If there are mixing motions in an atmosphere, then gas-phase condensible species can be be transported to regions where condensation can occur. In 1D models, the competing effects of upward mixing and downward transport due to sedimentation of the condensed phase must be treated to appropriately model vertical transport of condensible materials \citep{ackermanandmarley2001}. The following subsections describe a model that incorporates the key 1D physics of radiative energy transport, convective transport, and cloud microphysics.

\subsection{\texttt{CLIMA}: A 1D Radiative-Convective Terrestrial Planetary Climate Model}

We adopt \texttt{CLIMA}, a 1D (vertical) plane-parallel radiative-convective climate model that has been widely applied in the past to constrain the liquid water Habitable Zone  \citep{kastingetal1993,kopparapuetal2013}, and to explore the climates in Venus-\citep{kasting1988,vidaurrietal2022}, Mars-\citep{kastingetal1997}, and Earth-like \citep{kasting1988,pavlovetal2000} environments. More recently, \texttt{CLIMA} has been used to model and explore the climates of terrestrial exoplanets across a wide range of stellar and orbital configurations \citep{ramirezandkaltenegger2014} as a means of characterizing and exploring their radiative environments. Most applications of \texttt{CLIMA} omit cloud treatments and, instead, account for clouds by tuning the grey surface albedo until an Earth-like surface temperature is achieved for Earth-like model inputs \citep{kasting&ackerman1986,kasting1988,kastingetal1993,pavlovetal2000,kopparapuetal2013}. It should be noted here that adjusting the surface albedo of the planet may produce the correct converged surface temperature, but it does not produce realistic profiles of energy fluxes for an Earth-like atmosphere \citep{Kitzmann2010,Kitzmann2011,ZSOM2012,fouchezetal2018}. This is likely due to the neglect of the significant radiative effects clouds have on the overall atmospheric thermal and radiative structure.

\texttt{CLIMA} is generally used as either a \textit{forward model} or \textit{inverse model}, though both prescribe a convective troposphere upon model initialization. The \texttt{CLIMA} forward model fixes the top-of-atmosphere incident stellar flux and iteratively steps towards an equilibrium solution by computing and applying layer dependant radiative heating rates,
\begin{equation}
    \frac{\partial T}{\partial t}=\frac{g}{c_{\rm p}}\frac{\partial F}{ \partial p} \ ,
    \label{eqn:thermal_gradient}
\end{equation}
where $t$ is time, $F(p)$ is the net flux of radiative energy through each layer, $g$ is the altitude-dependent acceleration due to gravity, and $c_{\rm p}$ is the altitude-dependent heat capacity. Any model layers that achieve a super-adiabatic thermal gradient due to radiative energy exchanges are relaxed to a prescribed adiabatic structure. Model timestepping is not fully physical as only an equilibrium solution is desired\,---\,energy conservation can be omitted to speed up convergence and layers are assumed to convectively adjust instantaneously. Iterations are repeated and the atmospheric thermal structure is adjusted until the model reaches a desired level of convergence. Here, we define convergence of the forward model when the net thermal flux and net solar flux at the top of the atmosphere fall within 0.001\% of each other. 

The inverse version of \texttt{CLIMA}, established by \citet{kasting1988} and used in the Habitable Zone calculations of \citet{kastingetal1993} and \citet{kopparapuetal2013}, specifies a surface temperature and then prescribes a  temperature-pressure profile that assumes a convective troposphere with an overlaying isothermal stratosphere. Shortwave and longwave radiative flux profiles are computed for this adopted atmospheric model and \texttt{CLIMA} then determines the effective solar/stellar radiation necessary for maintenance of the specified atmospheric structure ($S$),
 \begin{equation}
     S_{\rm eff} \equiv \frac{S}{S_{\rm 0}} =\frac{F_{\rm t}}{F_{ \rm s}},
     \label{eqn:S_eff}
 \end{equation}
where $S_0$ is the solar constant ($S_{\rm 0}$ = 1,360\,W\,m$^{-2}$), $S_{\rm eff}$ is the effective flux $S$ relative to the solar flux at 1 au from the Sun, $F_{\rm t}$ is the top-of-atmosphere net thermal flux, and $F_{\rm s}$ is the top-of-atmosphere net solar/stellar flux. Here, $F_{\rm s}$ is determined using an incident solar flux of $S_0$ and this inverse approach only guarantees top-of-atmosphere planetary energy balance \citep{kastingetal1993}.

In the inverse climate model, and the initial timestep of the forward model, \texttt{CLIMA} computes the atmospheric thermal structure based on a surface temperature, a prescribed water vapor relative humidity profile, and a surface \ch{CO2} volume mixing ratio. Level temperatures are computed from the surface upwards following a prescribed adiabat until the temperature falls to the user-provided isothermal temperature in the stratosphere, thereby defining the initial radiative-convective boundary. In convective regions, the adopted adiabat can be a pure dry adiabat or a moist \ch{H2O} or \ch{CO2} pseudo-adiabat \citep[see Appendix A of][]{kasting1988}. Absent a cloud/condensible model (described later), the prescribed water vapor profile can adopt a fully-saturated profile (i.e., 100\% relative humidity), a Manabe-Wetherald fixed relative humidity profile \citep{manabe&wetherald1967}), or a Manabe-Wetherald fixed relative humidity profile with a fixed stratospheric water vapor content \citep{kasting&ackerman1986,kasting1988,kastingetal1993,kopparapuetal2013}. More complex condensible vapor profiles (and associated convective adiabats) are enabled when a cloud model is introduced.

After the initial timestep in the forward model and following an application of the step-dependent radiative heating/cooling rates, the convective state of a layer is determined by first performing a condensation check for \ch{CO2} (i.e., a comparison between the layer \ch{CO2} partial pressure and the \ch{CO2} saturation vapor pressure at the layer temperature) to determine if the moist \ch{CO2} adiabat should be adopted across the layer. If \ch{CO2} is not condensing, an analogous check for \ch{H2O} is performed, taking into account the prescribed relative humidity. Should either \ch{CO2} or \ch{H2O} condense in a layer, its mixing ratio is adjusted to the appropriate relative humidity (fully saturated for \ch{CO2}) and follows the user-specified profile for \ch{H2O}. If neither species is condensing, or if the user specified a pure dry adiabat, the layer temperature gradient is compared to the dry adiabat and set to the latter if unstable to convection.

\subsection{Radiative Transfer}

The radiative transfer in \texttt{CLIMA} is based on the two-stream radiative transfer solutions outlined in \citet{toonetal1989}. The shortwave (0.2--4.35\,\textmu m) uses a delta-quadrature two-stream approximation applied over 38 equally spaced spectral bins. \texttt{CLIMA} uses correlated-$k$ coefficients to parameterize shortwave absorption by \ch{CO2} and \ch{H2O} \citep{kopparapuetal2013}. These correlated-$k$ coefficients consist of 8-term \ch{CO2} $k$-coefficients from the HITRAN 2020 database \citep{HITRAN2020} while the \ch{H2O} $k$-coefficients use HITRAN 2020 at low atmospheric pressures and HITEMP 2020 \citep{HITEMP2020} for pressures greater than or equal to 0.1\,bar. Other spectrally active species handled by the shortwave radiative transfer subroutines include \ch{CH4}, \ch{O3}, \ch{NO2}, and \ch{O2}, where total layer optical depths are determined by summing over individual gas optical depths weighted by layer volume mixing ratios \citep{kopparapuetal2013}. In spectral bins with wavelength values less than 1\,\textmu m, Rayleigh scattering contributes to the wavelength dependent layer optical depth. Increasing the mixing ratios of atmospheric condensates (such as water vapor) can significantly change the Rayleigh scattering effect in an atmosphere \citep{pierrehumbert2011}. For example, in moist atmospheres the Rayleigh scattering coefficient can reach values up to 30\% lower than that of dry air \citep{kopparapuetal2013}. The current model computes Rayleigh scattering mixing ratio-weighted optical depths based on layer composition using data previously measured in \citet{vardavasandcarver1984}.

In the longwave, opacities for \ch{CO2}, \ch{H2O}, and other species are incorporated into 55 spectral intervals extending from 0--15,000\,cm$^{-1}$. Eight-term $k$-coefficients for \ch{CO2} and \ch{H2O} are calculated using line-width truncations at 500\,cm$^{-1}$ and 25\,cm$^{-1}$, respectively. The $k$-coefficients are calculated over a temperature grid of (100--600\,K) and a pressure grid of (10$^{-5}$--10$^2$\,bar) \citep{kopparapuetal2013}. A short truncation width is used for \ch{H2O} because it is overlaid with a pressure-induced absorption continuum, as described in \citet{ramirezetal2014}. Layer optical depths are computed by convolving $k$-coefficients within each spectral bin. For \ch{CO2} wing absorption in the 15\,\textmu m region, \texttt{CLIMA} uses the 4.3\,\textmu m \ch{CO2} band $\chi$ factors as a proxy \citep{perrin&hartmann1989,ramirezetal2017}.  For \ch{H2O}, the BPS continuum of \citet{paynterandramaswamy} is used in its entire range of validity (0--19,000\,cm$^{-1}$) \citep{kopparapuetal2013}. Carbon dioxide pressure-induced absorption is parameterized using methods outlined in \citet{gruszka&borysow1997,baranovetal2004,ramirezetal2017}. The longwave radiative transfer routines also handle pressure-induced absorption by \ch{H2}-\ch{H2}, \ch{O2}-\ch{O2},\ch{N2}-\ch{H2}, \ch{CO2}-\ch{H2}, and line absorption from \ch{O3},\ch{C2H6},\ch{CH4}, and \ch{N2O}.

\subsection{Treatment of Fractional Cloudiness}

Following the methods outlined in \citet{marleyetal2010} \citep[see also][]{morleyetal2016}, we compute separate radiative flux profiles in distinct cloudy and cloud-free columns. These columns have identical profiles of thermal structure and gas-phase composition, while the cloudy column contains additional opacity sources from condensed-phase materials. Cloudy-column extinction optical depth, single-scattering albedo, and scattering phase function are determined via a weighted combination of the clearsky and cloud optical properties. The overall net radiative flux used to derive heating and cooling rates is computed by summing the individual cloudy and clearsky net fluxes ($F_{\rm cld}$ and $F_{\rm clr}$, respectively), 
\begin{equation}
    F(p) = f_{\rm cld}F_{\rm cld}(p) + (1-f_{\rm cld})F_{\rm clr}(p) \ ,
    \label{eqn:fractional_cloudiness}
\end{equation}
where $f_{\rm cld}$ is a user-defined parameter that corresponds to the cloud fractional coverage. In the global average, this flux-weighting method captures the overall effects of a partially clouded atmosphere and enables a 1D climate model to incorporate cloud models of various physical complexity.

\subsection{\texttt{EddySed}: A Parameterized Cloud Microphysics Tool}

\begin{figure}[ht]
    \centering
    \includegraphics[width=\textwidth]{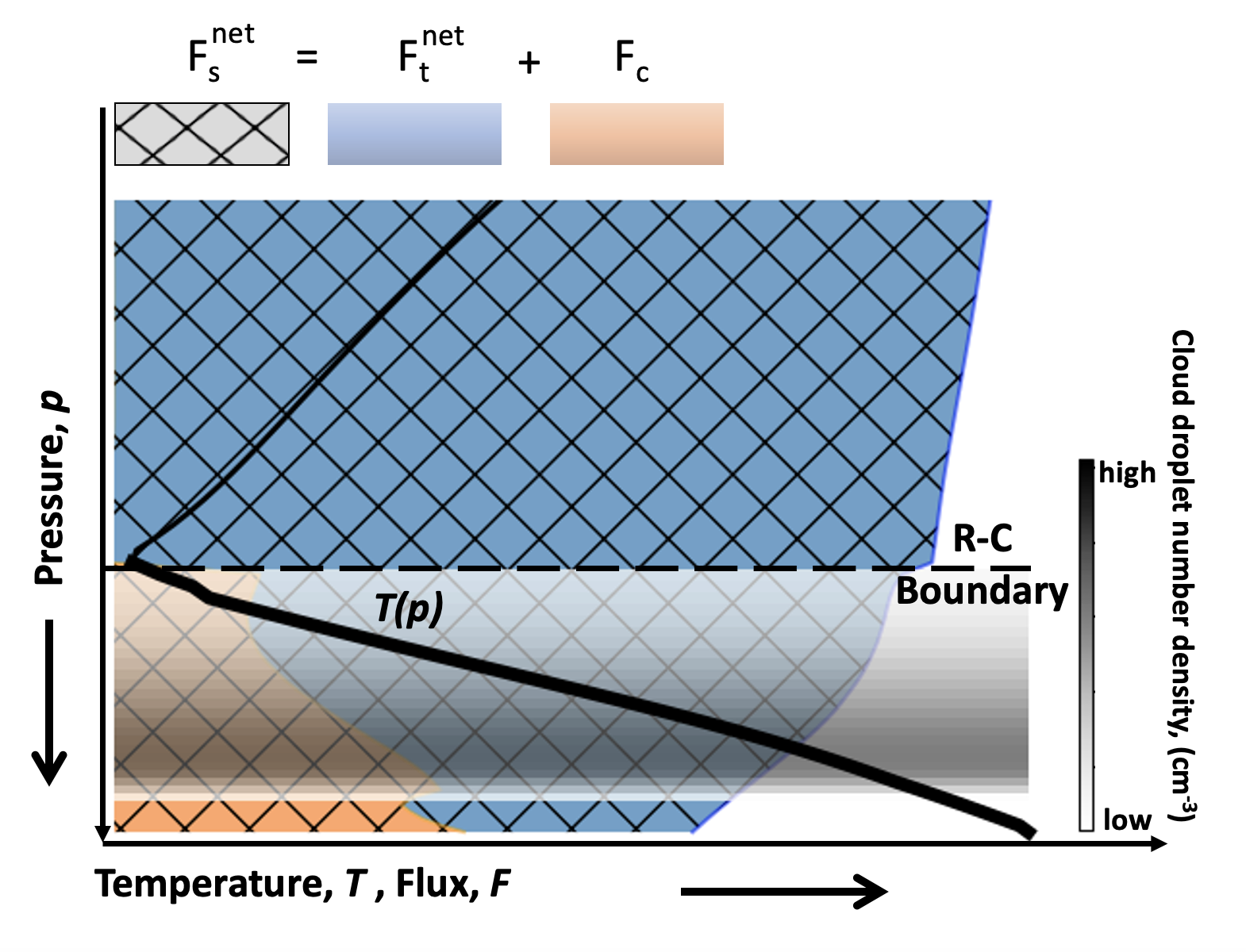}
    \caption{Model schematic. The vertical axis is pressure increasing downward, the horizontal axis shows relative temperature and energy flux, both increasing towards the right (away from the origin). Model levels are shown (horizontal, dashed line) and in this schematic model levels increment downward. The solid black line conveys the thermal structure of the atmosphere, where thickening corresponds to the convective troposphere. In the convective zone the model preforms a convective adjustment to ensure the thermal gradient remains that of a pseudo-moist adiabat. The gray box below the radiative convective boundary showcases where cloud formation takes place.  }
    \label{fig:model_cartoon}
\end{figure}

Clouds are modeled using the widely-adopted \texttt{EddySed} tool \citep{ackermanandmarley2001}, which provides a balance between microphysical treatments and computational efficiency. In the most simple terms, \texttt{EddySed} is a 1D Eulerian framework where turbulent vertical diffusion mixes moist air upward until it saturates and condenses. Condensed-phase material will rain out via sedimentation so that, in steady state, upward diffusion of material balances downward sedimentation \citep{ackermanandmarley2001}. Figure~\ref{fig:model_cartoon} is a cartoon demonstrating the overall atmospheric structure of \texttt{CLIMA} with a 1D cloud model like \texttt{EddySed}. 

In the \texttt{EddySed} model, the surface layer condensible mixing ratio is propagated upward while maintaining a constant molar mixing ratio below the cloud. Given decreasing temperature with pressure along the adiabat prescribed by \texttt{CLIMA}, the relative humidity increases with decreasing pressure until the condensation point. The deepest atmospheric layer to achieve saturation (and therefore condensation) is where the cloud model places the cloud base. Above the cloud base, vertical diffusion arising from atmospheric turbulence works to maintain a total mixing ratio ($q_{\rm t}=q_{\rm v}+q_{\rm c}$) that is scaled by the droplet sedimentation velocity and the convective velocity scale. Here, $q_{\rm v}$ is the molar mixing ratio of the condensate vapor phase (moles of vapor per moles of atmosphere) and $q_{\rm c}$ is the condensate mixing ratio (moles of condensate per mole of atmosphere). Below the atmospheric cold-trap, the condensing species vapor phase mixing ratios across both clearsky and cloudy columns are controlled entirely by \texttt{EddySed}. Adopting a uniform vapor phase mixing ratio across both columns prevents altitude-dependent pressure differences from forming between the cloudy column and clearsky subcolumns. Such pressure gradients would drive winds \citep{pierrehumbert1995,goldblatetal2013} that cannot be straightforwardly treated in a 1D tool. It should be noted here that this modeling scheme makes no attempt to address changes in condensible species vapor content below the cloud base \citep{ackermanandmarley2001} and, instead, relies on the atmospheric thermal structure to determine sub-cloud condensible species mixing ratios.

All condensation clouds in \texttt{EddySed} are modeled as horizontally homogeneous structures. The vertical extent of these structures and subsequent condensed phase molar mixing ratio and total molar mixing ratios of the condensible gas are computed in steady state by solving Equation~4 in \citet{ackermanandmarley2001} for each condensible species individually,
\begin{equation}
    -K_{\rm zz} \frac{\partial q_{\rm t}} {\partial z} - f_{\rm sed} w_{\rm *} q_{\rm c}=0 \ ,
\label{eqn:continuity}
\end{equation}
where $K_{\rm zz}$ is the vertical eddy-diffusion coefficient, $f_{\rm sed}$ is the ratio of the mass-weighted droplet sedimentation velocity to the convective velocity scale ($w_{\rm *}$), and $z$ is altitude. Equation~\ref{eqn:continuity}\,---\,the steady-state continuity equation\,---\,showcases that the upward mixing of the total condensible species (first term) is balanced by precipitation (second term). 

The average sedimentation velocity ($f_{\rm sed}w_{\rm *}$) offsets upward mixing from vertical advection and generally leads to $q_{\rm t}$ decreasing with altitude.  As in \citet{ackermanandmarley2001}, we adopt $f_{\rm sed}$ as a tuneable parameter which, together with $q_{\rm c}$, constrains how efficiently the cloud particles coagulate and ``sediment out'' of the condensing layers. The extreme case with no sedimentation to offset turbulent mixing ($f_{ \rm sed}=0$) is equivalent to ``frozen-in'' cloud structures that arise in some brown dwarf atmosphere models and represents cloud types in which there is no precipitation \citep{lunineetal1989,ackermanandmarley2001}.

We adopt the original eddy diffusion coefficient model from \citet{ackermanandmarley2001}, where $K_{\rm zz}$ is assumed to be the same as that for heat as derived for free convection \citep{gierasch&conrath1985}. Here, we use Equation (5) from \citet{ackermanandmarley2001},
\begin{equation}
    K_{\rm zz} = \frac{H}{3}\left(\frac{l}{H}\right)^{\frac{4}{3}}\left(\frac{R_{\rm U}F_{\rm c}}{\mu \rho_{\rm a} c_{\rm p}}\right)^{\frac{1}{3}} \ ,
    \label{eqn:K_zz}
\end{equation}
where $H$ is the atmospheric pressure scale height, $l$ is the turbulent mixing length, $R_{\rm U}$ is the universal gas constant, $\mu$ is the atmospheric molar weight, $\rho_{\rm a}$ is the atmospheric density, and $c_{\rm p}$ is the layer specific heat. All heat transport in the atmosphere is assumed to be fueled by a convective heat flux where, in convective regions of the atmosphere, the convective flux is the difference between the net shortwave and net longwave radiative fluxes (i.e., $F_{\rm c} = F_{\rm s}^{net} - F_{\rm t}^{net}$). The \texttt{EddySed} eddy diffusion coefficient scheme tends to overestimate values of $K_{\rm zz}$, but the adopted approach is consistent with all other well-tested applications of \texttt{EddySed} \citep{gaoetal2014a,Mang2022}. In purely radiative regions of the atmosphere (i.e., the thermal gradient is stable against convection) the mixing length-based prescription would yield a value of zero for $K_{\rm zz}$, which is unphysical. Thus, and instead of adopting a minimum eddy diffusion coefficient \citep[as in][]{ackermanandmarley2001}, we assign a minimum convective heat flux \citet{Mang2022}, 
\begin{equation}
     F_{\rm c}=h_{\rm c}F^{\rm{net}}_{\rm s}(p=0) \ ,
     \label{eqn:h_c}
\end{equation}
where $h_{\rm c}$ is a tuneable parameter that scales the minimum convective heat flux to the net flux at the top of the atmosphere ($F^{\rm net}_{\rm s}(p=0)$). We find that a minimum convective heat flux scheme with a value of 0.01 for $h_{\rm c}$ reproduces more realistic $K_{\rm zz}$ profiles in the upper atmosphere and prevents an abundance of cloud particles from forming directly above the radiative-convective boundary.

In general, for freely convecting atmospheres the mixing length is assumed to be roughly the pressure scale height. However, the free convection mixing length parameterization in \citet{ackermanandmarley2001} does not well capture what happens near a solid planetary surface. To capture a reduction in mixing length near a solid boundary, we adopt the Blackadar asymptotic mixing length \citep{blackadar1962}, with, 
\begin{equation}
    l=\frac{kz}{1 + kz/H} \ ,
    \label{mixing_length}
\end{equation}
where $k$ is the von K\'{a}rm\'{a}n constant \citep[we adopt a dimensionless value of $ 0.40$][]{hogstrom1988}. At large distances from the solid planetary surface ($z >> H$) the mixing length approaches $H$. For convective mixing close to the surface ($z < H$) the mixing length approaches $z$ \citep{ZHANG2021}. Note that some applications of the asymptotic mixing length introducing a scaling to the scale height \citep[e.g.,][]{lincowskietal2019}, which is not explored in the current work.

The cloud droplet size distribution is an important aerosol property and is directly affected by physical processes including upward mixing through evaporation, coagulation, and advection \citep{gaoetal2021}. In terrestrial clouds, precipitation is generally observed to increase in efficiency with increased mean particle size and increased spectral width \citep{ackermanandmarley2001,gaoetal2021}. Earth's liquid water convection clouds are generally bimodal in nature with condensation nodes peaking at a mean effective droplet radius of 10\,\textmu m accompanied by a larger precipitation mode \citep{ackermanetal2000b}. Precipitation efficiency can also be modified by additional factors such as the composition and distribution of cloud condensation nuclei \citep{pierrehumbert2011} and the magnitude of upwards convective mixing \citep{ackermanandmarley2001}. However, in general \texttt{EddySed} makes no assumptions about cloud microphysics beyond fitting a single layer-dependent log-normal size distribution  \citep[Equation~9 in][]{ackermanandmarley2001} to cloud particle sizes, and adopting a single value for $f_{\rm sed}$ throughout the cloud. Here,
\begin{equation}
    \frac{dn}{dr} = \frac{N}{r\sqrt{2\pi}\ln\sigma_{\rm g}}\exp\left[-\frac{\ln^{2}(r/r_{\rm g})}{2\ln^{2}\sigma_{\rm g}}\right] \ ,
    \label{eqn:number_dist}
\end{equation}
where $n$ is the number concentration of particles smaller than radius $r$, $\sigma_{\rm g}$ is the geometric standard deviation, $r_{\rm g}$ is the geometric mean radius, and $N$ is the total number concentration of particles. Adopting this same layer-dependent log-normal size distribution to our terrestrial cloud model allows cloud microphysics to be constrained with relatively few free parameters such as the standard geometric deviation from the geometric particle mean radius ($\sigma_{\rm g}$), the sedimentation efficiency ($f_{\rm sed}$), and $q_{\rm c}$ condensed phase mixing ratio. In turn, this allows \texttt{EddySed} to self-consistently compute the mean geometric particle radius ($r_{\rm g}$), the effective particle radius ($r_{\rm eff}$), the mass weighted sedimentation particle radius ($r_{\rm sed}$), and the total layer number concentration of particles.


Equation~10 from \citet{ackermanandmarley2001},
\begin{equation}
    f_{\rm sed} = \frac{\int_{0}^{\infty}v_{\rm f}(dm/dr)dr}{\epsilon \rho_{\rm a} w_{\rm *} q_{\rm c}} \ ,
\label{eqn:sedimentation_efficieny}
\end{equation}
defines the sedimentation efficiency, where $v_{\rm f}$ is the particle sedimentation velocity, $m$ is the particle mass, and $\epsilon$ is the ratio of condensate to atmospheric molecular weights.  In order to close the system analytically, \texttt{EddySed} finds the condensation cloud particle size that arises when we set the particle sedimentation velocity equal to the local convective velocity scale ($w_{\rm *} = K_{\rm zz}/l$). This approach reveals a power law relationship for particle sedimentation velocity, 
\begin{equation}
    v_{\rm f}=w_{\rm *} \left(\frac{r}{r_{\rm w}}\right)^{\alpha} \,
\label{eqn:powerlaw}
\end{equation}
where $\alpha$ is an exponent that is calculated from a fit to the particle fall speeds between $r_{\rm w}/\sigma$ and $r_{\rm w }$ when $f_{\rm rain}$ $\geq$ 1 and between $r_{\rm w}$ and $r_{\rm w} \sigma$ otherwise \citep[$\sigma$ is constrained to be $\geq$ 1.1 for the initial fit in][]{ackermanandmarley2001} and $r_{\rm w}$ is the mean upwelling particle radius.  In \texttt{EddySed} the particle sedimentation velocity for Stokes flow is computed by,
\begin{equation}
    v_{\rm f} = \frac{2}{9} \frac{\Delta \rho g r^{2} \beta}{\eta}.
\end{equation}
Here, $\Delta \rho$ is the difference in density between the condensing species and the background atmosphere, $g$ is the altitude-dependent gravitational acceleration, $\eta$ is the atmospheric dynamic viscosity, and $\beta = 1 +1.26 {\rm Kn}$ is the Cunningham slip correction number with,
\begin{equation}
    {\rm Kn} = \frac{1}{\sqrt{2}} \frac{k_{\rm B} T}{pr\pi d_{\rm m}^{2}} \ ,
    \label{eqn:knudson}
\end{equation}
the Knudsen number of the particle, $k_{\rm B}$ the Boltzmann constant, and $d_{\rm m}$ is the average weighted molecular diameter of the atmosphere. We further generalize \texttt{EddySed} through adopting an altitude-dependent depth of the Lennard-Jones potential ($\epsilon$) and atmospheric molecular diameter ($d_{\rm m}$) through linear combinations of data in \cite{rosner2000} and \texttt{CLIMA} layer mixing ratios. Resulting in, 
\begin{equation}
    \eta = \frac{5}{16} \frac{\sqrt{\pi \mu R_{\rm U} T}}{\pi d^2} \frac{\left( R_{\rm U} T / \epsilon \right)^{0.16}}{1.22}
    \label{eqn:eta}
\end{equation}
where $R_{\rm U}$ is the universal gas constant, and $\mu$ is the layer dependent atmospheric molecular weight.

The power law in Equation~\ref{eqn:powerlaw} allows Equation~\ref{eqn:sedimentation_efficieny} to be rearranged and integrated to reveal,
\begin{equation}
    r_{\rm g}=r_{\rm w} f_{\rm sed}^{1/\alpha} \exp\left(-\frac{\alpha + 6}{2}\ln^{2}\sigma_{\rm g}\right) \ ,
    \label{r_g}
\end{equation}
and,
\begin{equation}
    r_{\rm eff}=r_{\rm g} \exp\left(\frac{5}{2}\ln^{2}\sigma_{\rm g}\right) \ ,
    \label{r_eff}
\end{equation}
and,
\begin{equation}
    N = \frac{3\epsilon\rho_{\rm a} q_{\rm c}}{4\pi \rho_{\rm p} r_g^{3}}\exp\left(-\frac{9}{2}\ln^{2}\sigma_{\rm g}\right) \ ,
    \label{eqn:number_denisty}
\end{equation}
where we follow the prescription of \citet{ackermanandmarley2001} and take $\sigma_{\rm g}$ = 2 as the standard deviation of the log-normal fit to the condensation and coagulation modes of condensation cloud particle sizes. Constraining $f_{\rm sed}$ can illuminate relevant cloud microphysical properties by identifying the radius of mass-weighted sedimentation flux $r_{\rm sed} = f_{\rm sed}^{1/\alpha}r_{\rm w}$, leading to 
\begin{equation}
    f_{\rm sed} = \left(\frac{r_{\rm sed}}{r_{\rm w}}\right)^{\alpha}.
    \label{eqn:fsed_2}
\end{equation}

\subsection{Key Model Parameters}

The climate model is initialized though by specifying a few fixed parameters: the number of iterative timesteps, the initial dry surface pressure of the atmosphere, the initial global average surface temperature, the tropopause and upper atmosphere isothermal temperature, the initial surface layer relative humidity, the volume mixing ratios of noncondensible gases, the host star type, the stellar insolation, the grey planetary surface albedo the fractional cloudiness and the sedimentation efficiency. In the forward model, each iterative timestep self-consistently recomputes dynamic parameters such as the global average surface temperature, the global average thermal profile, the condensible species volume mixing ratios. Here, to ensure convergence of the dynamic parameters in the forward model, it is recommended to use 100+ iterative timesteps. Conversely, the inverse climate model diverges in setup through the number of specified model iterations. It is recommended that, for the inverse climate model, at least two iteration are used to ensure \texttt{EddySed} is stable.  



\section{Model Validation and Sensitivity}
\label{sec:Model Validation and Sensitivity}

We define an Earth-like model setup to enable validation tests of our patchy cloud climate model. This setup includes an Earth-like, normal-incidence insolation of 1,380\,W\,m$^{-2}$ from a Sun-like host and a background atmosphere of 78\% \ch{N2}, 21\% \ch{O2}, 360\,ppm \ch{CO2}, and \ch{Ar} as a filling gas. A grey surface albedo of 0.13 is adopted, which agrees with estimates of Earth's cloud-free Bond albedo \citep{Kitzmann2010,ZSOM2012}. Forward model runs are iterated for a burn-in period of 450 steps to help ensure convergence. The sections below explore sensitivity to the number of vertical model layers and, then, perform validation tests against Earth for a range of input parameters.

\subsection{Model Layer Tests}

\begin{figure}[ht]
    \centering
    \includegraphics{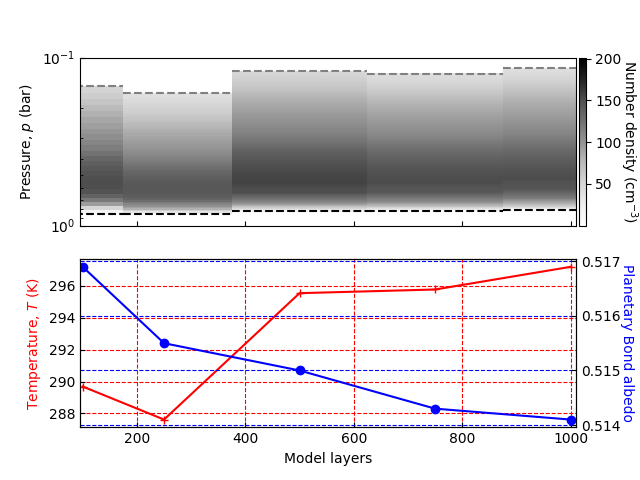}
    \caption{\textbf{Top:} Cloud layer number density as a function of model layers. The grey shaded regions correspond to the cloud condensate particle number densities. In each cloud structure, the black dashed lines correspond to where the cloud cumulative optical depths reach unity, integrated downward. Cloud particle number densities follow a similar structure regardless of the number of model layers.
    \textbf{Bottom:} Surface temperature (red; left vertical axis) and planetary Bond albedo (blue; right vertical axis) as a function of the number of model layers. The horizontal axis corresponds to the number of layers the climate model uses. Planetary surface temperature and Bond albedo stabilize after about 500 model layers.}
    \label{fig:layer_test_zsom_combined}
\end{figure}

Earlier, more heavily-parameterized 1D cloudy climate tools were shown to have sensitivity to the number of model layers below some critical resolution threshold \citep{ZSOM2012}. Thus, we test a grid of different log-pressure spaced models with 100, 250, 500, 750, and 1,000 layers. Models with more than 1,000 layers take a prohibitive amount of computational time to run. For testing purposes, we fix the fractional cloudiness at 50\% and adopt a sedimentation efficiency of $f_{\rm sed} = 1.0$. Models adopt an Earth-like surface relative humidity of 80\% with the stratospheric water vapor mixing ratio set by its value at the cold trap.

Figure~\ref{fig:layer_test_zsom_combined} demonstrates how planetary Bond albedo, surface temperature, and the cloud particle number density profile change with the number of prescribed model layers. Variations in planetary Bond albedo are relatively stable, with a maximum differential value of 3\%, while the planetary surface temperatures tend to stabilize after 500 model layers. Similarly, cloud particle layer number densities tend to stabilize after 500 layers, in good agreement with where the resulting global mean surface temperatures also stabilize. \citet{ZSOM2012} report similar model surface temperature stability after 350 vertical model layers. We use the results of this test moving forward and adopt a standard model vertical resolution of 500 layers.

\subsection{Fractional Cloudiness Tests}
\begin{figure}[ht]
    \centering
    \includegraphics{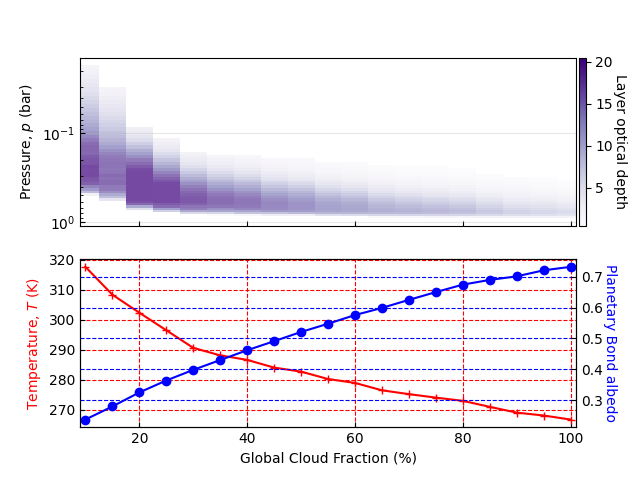}
    \caption{\textbf{Top:} Cloud vertical extent as a function of fractional cloudiness. Cloud optical depth is indicated by the depth of the purple color, with darker purple corresponding to layer optical depth that is large, whereas lighter colors correspond to layer optical depth that is small. Layer optical depths above 20 are not resolved to better enable the depiction of optically thinner cloud layers.  
    \textbf{Bottom:} Planetary surface temperature (red; left Y-axis) and planetary Bond albedo (blue; right Y-axis) as a function of the fractional cloudiness. Planetary surface temperature is highest for a low fractional cloudiness of 10\% and reaches a minimum for a fractional cloudiness of 100\%. Planetary Bond albedo reaches a minimum at the low fractional cloudiness of 10\%, and reaches a maximum at a fractional cloudiness of 100\%. An Earth-like Bond albedo of 0.33 corresponds to a surface temperature of 273 K and a fractional cloudiness of 20\%. }
    \label{fig:fcloud_test_zsom_combined}
\end{figure}

Fractional cloudiness is a critical model parameter that cannot be predicted in a 1D framework. To demonstrate sensitivity to this parameter, we explore our standard Earth setup with fractional cloudiness varying from 5\%--100\%. A 0\% cloudy (i.e., fully clearsky) case is omitted as the adopted low surface albedo tips simulations into a runaway state. As before, $f_{\rm sed}$ is assigned a value of 1.0 to yield mean particle radii close to those reported for Earth clouds \citep{ackermanandmarley2001}.

Figure~\ref{fig:fcloud_test_zsom_combined} shows how cloud thickness and extent, surface temperature, and Bond albedo vary with prescribed cloud fraction. Planetary Bond albedo increases linearly with cloud fraction. This relationship occurs because cloud reflectivity is large across all cloud fractions and is consistent with Earth cloud-cover measurements \citep{kingetal2013,Katoetal2020,Proud2021}.
Similarly, the reflective clouds lead to high Bond albedos at larger cloud fractions where the resulting high planetary reflectivities cause an overall cooling of surface temperatures. Realistic values for Earth's global-average surface temperature and Bond albedo are 288.5\,K and 30\%, respectively \citep{lacis2012,mallamaetal2017}. At a cloud fraction of 40\% our model yields a surface temperature of 287\,K and Bond albedo of 46\% whereas a cloud fraction of 20\% gives a more Earth-like Bond albedo of 33\% but a higher surface temperature at 301\,K. Thus, our model is capable of producing reasonably Earth-like climate states, but matching both Earth's surface temperature and Bond albedo would require either more detailed treatments of clouds or a 3D tool.

Figure~\ref{fig:fcloud_test_zsom_combined} also explores the effect of cloud fraction on the overall cloud structure. In general, cloud structures are a function of cloud feedbacks on the atmospheric radiative profiles and the resulting overall thermal structures. Most apparently cases with smaller cloud coverages (i.e. 10-25\%) produce more vertically-extended cloud structures, which is suggestive of an abundance of convective mixing and higher absolute humidity in these warmer conditions. Because of this, warmer climates tend to have vertically extended tropospheres when compared to cooler climates. Similarly, warmer climates push the height of the cloud base upwards because these cases have a reduced deep-atmosphere temperature gradient, meaning the vapor phase of the condensible species must travel further along the adiabat to condense. \citet{kingetal2013} \citep[as well as][]{Katoetal2020,Proud2021} report global fractional cloudiness values of 60\%--70\%, whereas our climate model tends to result in Earth-like global average surface temperatures for fractional cloud values of 30-45\%. Here, measured global fractional cloudiness values from the literature \citep{kingetal2013,Katoetal2020,Proud2021} account for the presence of thin cirrus clouds that have a markedly smaller impact on planetary Bond albedo. \texttt{EddySed} clouds are representative of deep-atmosphere clouds and are in good agreement with the 30-50\% of Earth's deep-atmosphere cloud cover \citep{kingetal2013,Katoetal2020,Proud2021}.

\subsection{Surface Relative Humidity Tests}

\begin{figure}[ht]
    \centering
    \includegraphics{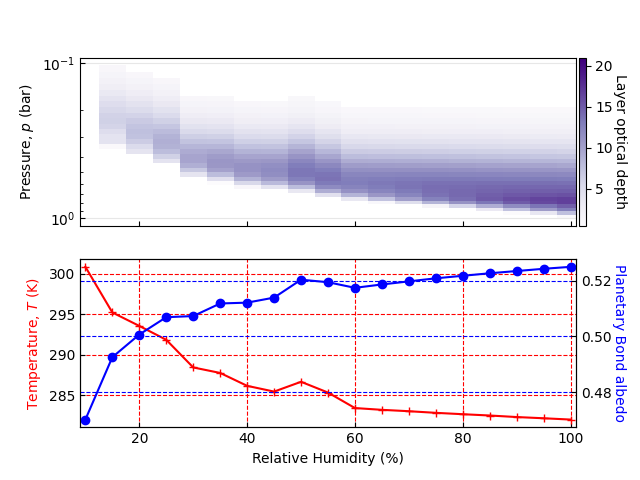}
    \caption{\textbf{Top:} Cloud vertical structure as a function of global surface relative humidity. Shading indicates layer optical depth with darker purple shades corresponding to more-opaque layers. 
    \textbf{Bottom:} Global surface layer temperature (red; left Y-axis) and planetary Bond albedo (blue; right Y-axis) as a function of global surface relative humidity. 
    Planetary surface temperature reaches a maximum at 301 K and a minimum at 282 K. The planetary Bond albedo spans a range of 0.47 to 0.525.}
    \label{fig:relhum_test_zsom_combined}
\end{figure}

Tropospheric water vapor content has a significant effect on the overall atmospheric thermal structure and resulting cloud profiles. To explore sensitivity to our humidity prescription, we vary the surface relative humidity parameter from 0\% to 100\% in increments of 5\%. Fractional cloudiness and sedimentation efficiency are fixed at 50\% and 1.0, respectively. At a surface relative humidity of 0\% the model defaults to a minimum water vapor mixing ratio of $4\times10^{-8}$. Figure~\ref{fig:relhum_test_zsom_combined} demonstrates how cloud extent, surface temperature, and Bond albedo respond to changes in surface relative humidity.

At low surface relative humidity, the small condensate species vapor phase mixing ratio must be propagated along the adiabat to markedly lower atmospheric pressures for condensation to occur as compared to cases with large surface relative humidity. It should be noted here that the cloud tops are also significantly higher in the atmosphere, likely due to the gradual increase of the eddy diffusivity causing a significant increase in upward mixing. This is demonstrated in Figure~\ref{fig:relhum_test_zsom_combined}, where for low overall relative humidity the cloud decks increase in altitude until vanishing completely at a relative surface humidity of 0\%. At above 60\% surface layer relative humidity, the cloud top altitudes tend to stabilize and the overall cloud structures generally resemble one another. Here, the major difference arises where the cloud model places the first layer of condensation. 

It is reasonable to interpret these results as a way of controlling cloud base placement, which can have some effect on the overall planetary Bond albedo and planet-wide mean surface temperatures. Most notably, the surface temperature of the planet tends to stabilize after a relative surface humidity of 60\% and drop in a linear fashion from roughly 284\,K to 282\,K. Similarly, the planetary Bond albedo tends to increase in an oppositely-sloped trend from roughly 0.52 to 0.53 from 60\% relative surface humidity to 100\% relative surface humidity. This effect is markedly less than the effects of overall cloud cover but nonetheless has important implications for the radiative environments inside of the climate model and does have a significant effect the atmospheric thermal structure. For true Earth-like climate models we suggest adopting a surface relative humidity of around 70-80\%, as these are representative of Earth's global average relative humidity \citep{lacis2012} and result in realistic Earth-like cloud base altitudes \citep{Katoetal2020,Proud2021}. 

\subsection{Earth-like Case Validations}

Informed by the experiments above, we adopt a baseline Earth-like model for validation purposes that has a global surface relative humidity of 70\% \citep{manabe&wetherald1967}, a fractional cloudiness of 55\%, and a sedimentation efficiency of $f_{\rm sed}=0.6$. This sedimentation efficiency is slightly smaller than the value used above but, importantly, agrees with derived values relevant to Earth stratocumulus clouds \citep{ackermanandmarley2001} and yields droplet geometric radii on the order of 11--13\,\textmu m and cloud droplet effective radii ranging from 40--50 \textmu m. With these adopted parameters, the model produces an Earth-like surface temperature of 292\,K and a deep troposphere undergoing moist convection.

\begin{figure}[ht]
    \centering
    \includegraphics{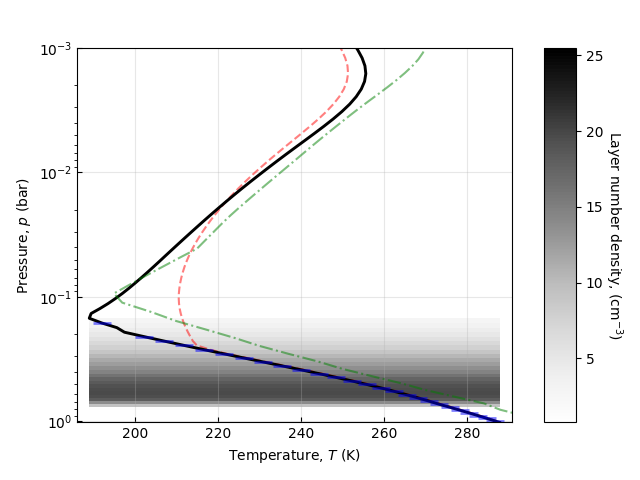}
    \caption{Cloud and atmospheric thermal profiles of the baseline Earth-like model (black solid line) as compared to a ICRCCM mid-latitude Earth sounding (green dashed-dot line), and the cloud free Tuned \texttt{CLIMA} model (red dashed line). In the troposphere, model moist convection is flagged with a translucent horizontal blue line at each differential layer boundary. The tropopause temperature in this model case is 1901\,K, and the surface global average temperature is 292\,K. The model tropospheric thermal gradient is shallower than that of the ICRCCM mid-latitude sounding.}
    \label{fig:Best_Earth-like_Test_Case_100layers}
\end{figure}

Figure~\ref{fig:Best_Earth-like_Test_Case_100layers} compares our baseline Earth-like model thermal structure to a representative thermal structure for Earth \citep[adopted by the InterComparison of Radiation Codes used in Climate
Models {[ICRCCM]} study;][]{mcclatcheyetal1972,lutheretal1988,ellingsonetal1990}. The representative profile is typical of Earth's tropics and is selected as the cloud structures produced in our simulations are most analogous to convective clouds formed in Earth's equatorial regions. Also shown is a ``Modern Earth'' template case from the tuned, cloud-free \texttt{CLIMA} tool. Overall the tropospheric thermal gradients are comparable in each model run. At the tropopause, each atmospheric thermal profile tends to disagree. The tropospheric temperature for the cloudy Earth-like model is 191\,K, whereas the tropospheric temperature for the tropical ICRCCM sounding tropopause is 195\,K \citep{lutheretal1988,ellingsonetal1990}. Likely, the radiative heating rate is diminished above the cloud layer. \citet{Randalletal1989} \citep[and][]{Katoetal2020} show with the UCLA/GLA general circulation model that clouds warm the tropical atmosphere below 7,000\,m and cool near the tropopause. Further, deeply convective cloud structures, such as the clouds modeled by \texttt{EddySed}, could be mimicking a convective overshooting top in which the convective energy of the troposphere carries condensates above the tropopause and cools the surrounding lower stratosphere \citep{Katoetal2020,Proud2021}. The resulting tropopause temperatures in these scenarios range from 180-220\,K and our tropospheric temperatures are well within these ranges.

The only regions in our baseline Earth-like model that are completely saturated are the atmospheric layers immediately adjacent to the cloud. This results in a smaller amount of non-condensing water vapor below the cloud deck when compared to model cases that are completely saturated and results in a net greenhouse effect of 250\,W\,m$^{-2}$, which we compute by comparing the net upwelling fluxes at the planetary surface to the net upwellling fluxes at the top of the atmosphere. However, this overshoots the value reported by \citet{lacis2012} of roughly 150\,W\,m$^{-2}$. This likely arises due to water vapor feedbacks that can't be accounted for in 1D models, such as moist-upwelling convective plumes and dry downwelling plumes that serve as ``radiator fins'' in tropical regions on the Earth \citep{pierrehumbert1995} in conjunction with cloud greenhouse effects. Notably, the resulting global average thermal profiles are representative of Earth due to the somewhat high planetary Bond albedo (here, 0.54) which results in a decrease of the absorbed incident radiation and balances the enhanced greenhouse effect.

\begin{figure}[ht]
    \centering
    \includegraphics{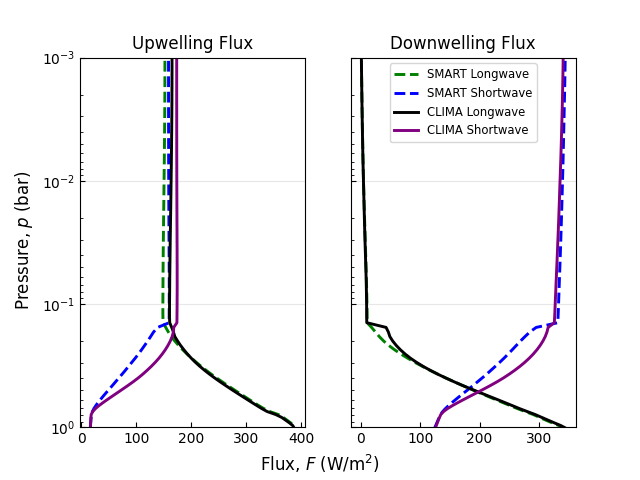}
    \caption{Upwelling (left) and downwelling (right) radiative fluxes for \texttt{CLIMA} (solid) and SMART (dashed). Green lines correspond to thermal fluxes and black lines correspond to solar fluxes. Deep atmospheric flux profiles are in good agreement.}
    \label{fig:Net_Fluxes_new}
\end{figure}

Our baseline Earthlike case also presents an opportunity to validate \texttt{CLIMA}'s radiative transfer routines, which is especially important as these routines have not seen extensive applications to cloudy atmospheres. To do this, we compare net radiative fluxes computed by \texttt{CLIMA} to those computed by the more-sophisticated Spectral Mapping Atmospheric Radiative Transfer (\texttt{SMART}) model \citep{meadows&crisp1996,robinson2017} given identical atmospheric structure inputs. Figure~\ref{fig:Net_Fluxes_new} showcases upwelling and downwelling flux profiles for the baseline Earth-like case from \texttt{CLIMA} and from a 32-stream \texttt{SMART} simulation. After assigning a gray Lambertian planetary surface in \texttt{SMART} we perform a direct comparison by ingesting an atmospheric structure profile from the converged baseline Earth-like \texttt{CLIMA} model. Promisingly, \texttt{CLIMA}'s Delta-Eddington radiative transfer subroutines produce comparable flux profiles to the more computationally-expensive \texttt{SMART} results. However, according to \citet{hengetal2018} the original two-stream solutions described in \citet{toonetal1989} do not preform well in the presence of large aerosol radiative scattering (such as those arising in condensate clouds) due to the overestimation of forward-scattered radiation from the adopted Dirac delta-scaled scattering phase function. Such an overestimation would explain the larger downwelling shortwave fluxes through the upper regions of the cloud in the \texttt{CLIMA} model versus the \texttt{SMART} case.

\section{Results}
\label{sec:results}
The global average fractional cloudiness and sedimentation efficiency for terrestrial worlds remain uncertain. We account for this uncertainty by first exploring sensitivity to a wide range of potential cloud properties for Earth-analog climates and then down-select to a series of Earth-like test cases which are directly compared to other Earth spectral models. Further, we create a series of super-Earth models following a similar framework and generate a series of coupled climate-to-spectral models.

\subsection{Exploring Cloud Parameters}
\label{subsec:Exploring Cloud Parameters}

For any detected Earth-like exoplanet, the global fractional cloudiness and a representative sedimentation efficiency would be uncertain. Some preliminary studies applying \texttt{EddySed} to marine stratoculumus over the pacific have somewhat constrained the parameter space needed for terrestrial water cloud representation \citep{ackermanetal2000b,ackermanandmarley2001}, but adopting these values for global cloud properties on exoplanets cannot be easily justified. To understand the variety of Earth-like worlds that could exist for a range of cloud scenarios, we explore a grid of model climates with fractional cloudiness ranging from 0 to 100\% and a sedimentation efficiency ($f_{\rm sed}$) ranging from a value of 0.01 to a somewhat efficient value of 3.0. The exploration is performed for a Sun-like host with the climate model in its ``inverse'' mode, which then determines the insolation (or, equivalently, orbital distance) where the particular cloud setup would produce a surface temperature of 288.5\,K. Surface relative humidity is fixed at 100\% (which produces a maximum amount of water vapor greenhouse warming) as there is no simple motivation for adopting Earth's known global-average surface relative humidity or any other particular value. For inverse calculations, the stratospheric isotherm is set to 200\,K, and testing indicated little sensitivity to this parameter. Finally, we adopt an atmospheric composition with modern Earth-like levels of \ch{N2}, \ch{O2}, \ch{O3}, \ch{CO2}, and \ch{Ar}.

\begin{figure}[ht]
    \centering
    \includegraphics{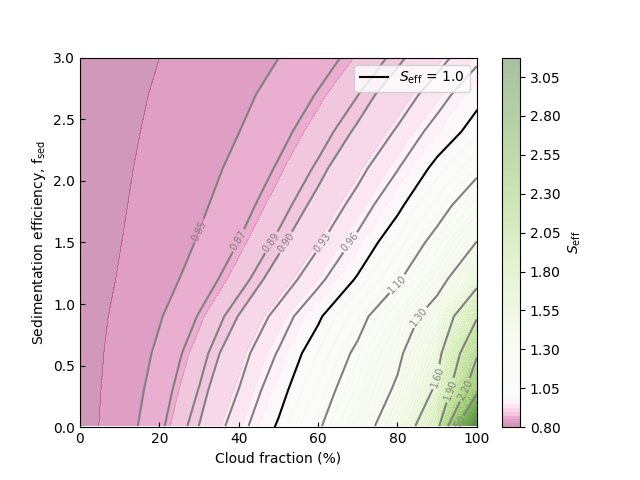}
    \caption{Effective insolation required for planetary energy balance ($S_{\rm eff}$; contours) as a function of the cloud fraction (horizontal axis) and sedimentation efficiency (vertical axis). A $S_{\rm eff}$ value of unity corresponds to a case where the climate state could be maintained with Earth's current level of insolation.}
    \label{fig:Seff_paper}
\end{figure}

Figure~\ref{fig:Seff_paper} demonstrates the resulting value of $S_{\rm eff}$ for each of the climate states as a function of global cloud fraction and sedimentation efficiency. Note that $S_{\rm eff}$ values above or below unity indicate which combinations of cloud fraction and sedimentation efficiency have a net cooling versus net warming effect on the overall planetary climate, respectively. Where $S_{\rm eff}$ = 1.0, maintaining top-of-atmosphere planetary energy balance requires the same amount of incident stellar flux as the Earth receives. Where $S_{\rm eff} < 1$, the particular combination of cloud and atmospheric parameters lead to a net warming effect and planetary energy balance would be achieved at insolations smaller than what Earth receives. Conversely, where $S_{\rm eff} > 1$ the solar flux incident on the planet must be increased to overcome a net cooling effect that results from the combination of cloud and atmospheric parameters. The majority of the explored parameter space shows a net warming effect on the overall planetary climate. This likely arises due to the large water vapor greenhouse effect that stems from the assumption of a fully-saturated atmosphere at the surface. Evidence for this interpretation also comes from the cloud-free models (i.e., 0\% cloud fraction) having $S_{\rm eff} < 1$.

\begin{figure}[ht]
    \centering
    \includegraphics{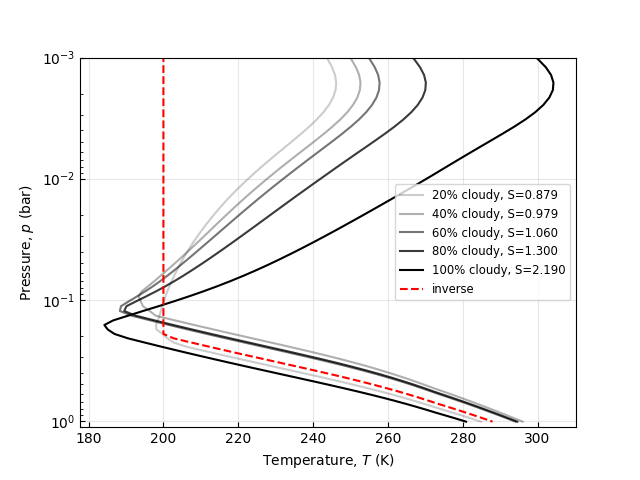}
    \caption{Atmospheric thermal profiles from the forward climate model with various values of insolation (grey shades) versus a profile from the inverse climate model (red). For the forward models, the insolation is guided by results in Figure~\ref{fig:Seff_paper} so that the equilibrium thermal structure should be similar to the inverse model. Ozone warming is in the stratopause is driven by the incident amount of flux on the top of the planetary atmosphere. In this set of forward models there is a divergence in the stratopause temperature due to changes in the amount of incident flux (i.e. S=0.879 results in less ozone warming above the cloud deck, whereas S=2.190 results in large ozone warming above the cloud deck). Overall the surface temperatures in each forward case are near to that provided by the inverse climate model.}
    \label{fig:inverse_informed_thermalprof}
\end{figure}

Results in Figure~\ref{fig:Seff_paper} indicate a striking range of insolations that the inverse climate model predicts to be consistent with a 288.5\,K surface temperature. Thus we select a number of scenarios to explore in more detail with the \textit{forward} model that span a range of fractional cloudiness and at a fixed sedimentation efficiency where adopting the insolation from Figure~\ref{fig:Seff_paper} should yield a surface temperature that is roughly consistent with the adopted structure in the \textit{inverse model}. For a sedimentation efficiency 0.5 and fractional cloudiness values of 20\%, 40\%, 60\%, 80\%, and 100\%, Figure~\ref{fig:inverse_informed_thermalprof} showcases the resulting forward model thermal profiles. In short, this experiment quantifies how well the inverse model performs when predicting climate states from top-of-atmosphere energy balance arguments.

Another striking aspect of Figure~\ref{fig:Seff_paper} is the wide range of cloud parameters that would yield an Earth-like climate at Earth's true insolation. For example, a sedimentation efficiency of 0.5 and a fractional cloudiness value of 54.8\% and a fractional cloudiness value of 100\% with a sedimentation efficiency of 2.57 both lie along the $S_{\rm eff}=1.0$ curve. Here, we use the forward climate model to investigate four potentially Earth-like climate states around this curve. Cloud parameters for these four cases are given in Table~\ref{tab:earthlike} where resulting temperature and albedo values are then compared to the cloud-free version of \texttt{CLIMA} (where tuning the surface albedo is required to yield an Earth-like surface temperature).

\begin{figure}[ht]
    \centering
    \includegraphics{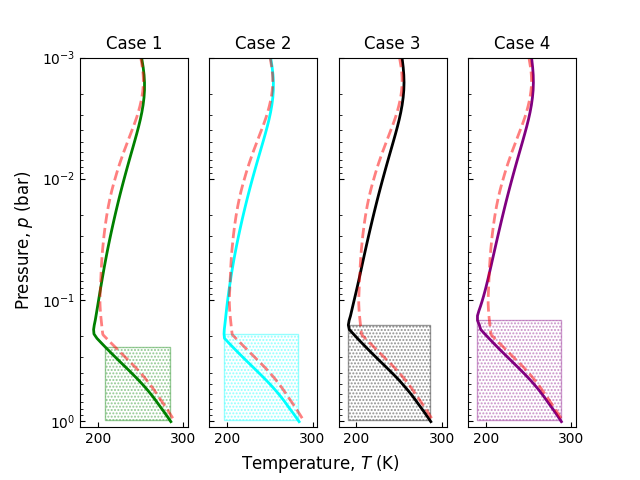}
    \caption{Atmospheric thermal structures for highlighted Earth-like test cases (solid colors) and the thermal structure from clearsky \texttt{CLIMA} where the surface albedo is tuned as in previous works (red dashed). Cloud vertical extent is represented by background shaded boxes. Overall the cloudy climate models have colder tropopause temperatures than the tuned clearsky model but have similar stratopause and surface temperatures.}
    \label{fig:thermal_structure_Earth-like_combine}
\end{figure}

Figure~\ref{fig:thermal_structure_Earth-like_combine} shows thermal and cloud structures for the four adopted cases, demonstrating that a large range of potential cloud properties can yield a roughly Earth-like thermal structure. We find that, despite resulting Earth-like climates, our atmospheric thermal profiles differ slightly from that as computed by the tuned cloud-free \texttt{CLIMA} case. Each slight difference in atmospheric thermal structure likely arises due to different and competing cloud radiative effects in each of the unique climate states. For instance, the 100\% cloudy case has a sedimentation efficiency of 3.0 which results in relatively efficient cloud droplet coagulation and subsequent cloud droplet particle growth to effective particle radii on the order of 130\,\textmu m and layer differential cloud optical depths averaging around 2.5. Subsequently, the cloud top pressure in Case~1 is deeper in the atmosphere (0.24\,bar) when compared to other cases. In each of these atmospheres, the climate model independently converges on thermal profiles that are close to each other suggesting that the sedimentation efficiency and fractional cloudiness parameters modulate the cloud greenhouse effect enough to maintain the desired globally averaged surface temperature of 288.5\,K. Ultimately, Case~1 results in optically thinner clouds\,---\,with large particle radii\,---\,and is representative of most climate cases with relatively high sedimentation efficiency and fractional cloudiness values. In comparison to this, Case~4 has a low sedimentation efficiency of 0.01 and a fractional cloudiness of 50\%, resulting in differential layer cloud optical depths of 220 and effective particle sizes near 12\,\textmu m (indicating cloud structures that do not efficiently precipitate out).

\begin{table}[ht]
\centering

\begin{tabular}{llllll}
\hline
Parameter & Case 1 & Case 2 & Case 3 & Case 4 & Tuned \texttt{CLIMA} \\ \hline
        Cloud fraction, $f_{\rm cld}$         &     100\%      &      75\%     & 60\% &      50\%     & n/a  \\ 
        Sedimentation efficiency, $f_{\rm sed}$           &       3.0      &      1.75     & 1.0  &      0.1      &  n/a  \\ 
        Surface Temperature (K)     & 286          & 284         & 287 & 288        & 290 \\ 
        Effective Temperature (K)   & 219          & 222         & 226  & 232       & 263 \\ 
        Tropopause Temperature (K)  & 195          & 196         & 190 & 190        & 203 \\
        Stratopause Temperature (K) & 255          & 254         & 256 & 256        & 253 \\ 
        Bond Albedo             & 0.63           & 0.59          & 0.56 & 0.52          & 0.21 \\
\end{tabular}
\caption{Earth-like Comparison Cases}
\label{tab:earthlike}
\end{table}

\begin{figure}[ht]
    \centering
    \includegraphics{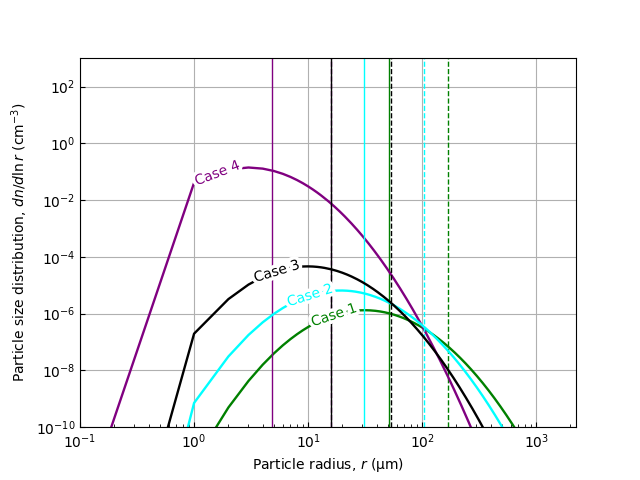}
    \caption{Log-normal cloud droplet number density at the cloud mid-point for Case~1 (green), Case~2 (cyan), Case~3 (black), Case~4 (purple). The dashed horizontal lines represent the mean effective cloud droplet radii. The solid horizontal lines correspond to the mean geometric cloud droplet radii. Overall, climates with higher sedimentation efficiency result in cloud droplet number densities that are diminished, but have larger cloud droplet radii. }
    \label{fig:particle_radii_combined}
\end{figure}

Figure~\ref{fig:particle_radii_combined} further compares cloud particle sizes across our four identified Earthlike cases. Here we show the droplet distributions at the midpoint of the cloud in each Earthlike case. Here, the numerical cloud midpoint is found by isolating the log pressure spaced layers that have cloud optical properties (i.e., where clouds are present) and determining the midpoint in pressure space between these layers. Figure~\ref{fig:particle_radii_combined} showcases the cloud droplet size distributions of the cloud layer midpoints. For each cloud midpoint a numerical ``slice'' is taken as a window into the cloud particle distributions. We also plot, as vertical lines, the effective particle radii (dashed lines) and the geometric particle radii (solid lines). In general, there is a trend corresponding to higher number densities and smaller geometric and effective cloud particle radii, that is inversely proportional to the initial value of $f_{\rm sed}$. In clouds with low sedimentation efficiency, cloud particles do not grow to large particle radii and are representative of atmospheric environments with large abundances of cloud condensation nuclei \citep{pierrehumbert2011}. The resulting number densities and cloud particle sizes are ultimately what drive the cloud optical properties. 

In Case 1, where the fractional cloudiness is 100\%, the total planetary Bond albedo reaches a value of 0.63, which is more than double that of the true Earth at 0.30 \citep{lacis2012,mallamaetal2017}. In this case, the radiative effect of the 100\% cloud cover should have a drastic cooling effect on the planetary atmosphere, but the globally averaged surface temperature is 286 K. In contrast to this case, the Tuned \texttt{CLIMA} template has a fractional cloud cover of 0\% and has a planetary Bond albedo of 0.21 \citep{kopparapuetal2013}, significantly lower than the value measured in \citep{mallamaetal2017} with a surface temperature of 290 K. This implies that for the climate in Case 1, the clouds not only introduce a higher planetary Bond albedo, but also produce a strong greenhouse effect. This is similar to the planet Venus, in which despite a high planetary Bond albedo of 0.75 \citep{bullocketal2017} a significant greenhouse flux is driven by CO2 absorption leading to global average surface temperatures of 740~K \citep{bullock&grinspoon2001,bullocketal2017}. Similarly, in our cloudy climate models, the combined greenhouse effects from water vapor and water clouds result in a clement surface despite a high planetary Bond albedo.

\subsection{Spectra of Earth-like Cases}

\begin{figure}[ht]
    \centering
    \includegraphics{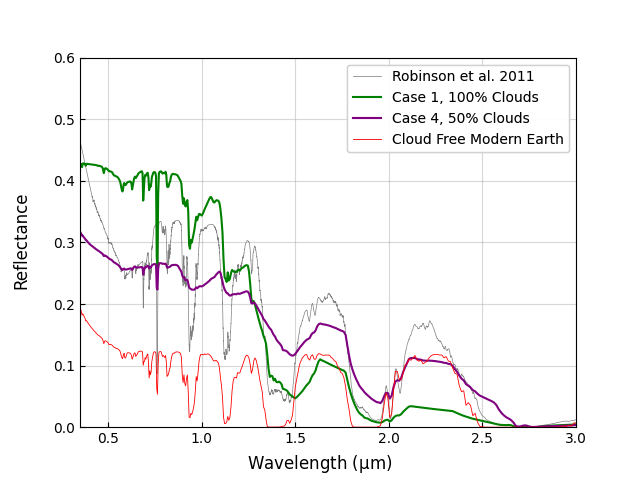}
    \caption{Planetary reflectance as a function of wavelength for Earth-like example cases around a Sun-like host star. As shown in Table~\ref{tab:earthlike}, cloud fraction decreases as case number increases. We also show a high-fidelity 3D spectral simulation of Earth for comparison \citep{robinsonetal2011}. All model reflection spectra showcase high reflectivity at visible and shorter near-infrared wavelengths.  At longer near-infrared wavelengths, overall larger droplets in lower case-number simulations result in more-absorptive clouds, thereby flipping the relationship between continuum reflectivity and case number seen at shorter wavelengths.}
    \label{fig:Best_Earth-like_Test_Case_no_labels}
\end{figure}

Our selection of models with near identical thermal structures, yet very distinct cloud distributions, provides an opportunity to investigate the spectral appearance of several ``Earth-like'' worlds. Here, different cloud properties not only have a crucial effect on the overall radiative energy balance of the planetary atmosphere, but also have a significant effect on the overall emitted flux and reflected light spectrum of the planet. Due to the extreme differences in cloud cover percentage and overall effective cloud droplet sizes within each of these four climate models, it is useful to explore their effects on the overall reflected light spectrum. We independently craft a pipeline to port partially-clouded climate output models to the previously-mentioned SMART tool and use this to generate high-resolution synthetic spectra that we then degrade to a resolving power of 200. Figure~\ref{fig:Best_Earth-like_Test_Case_no_labels} showcases the resulting reflected-light spectra of our four Earth-like cases. In general, higher fractional cloudiness results in higher reflectivity shortward of  1.5\,\textmu m due to high cloud particle single-scattering albedos in this wavelength range. 

Inspection of molecular absorption features in the planetary spherical albedo spectra unveil a trend in which higher cloud fractions mute water vapor spectral feature absorption depth due to the diminished water vapor content in the upper atmosphere and the subsequent diminished fractional deep atmosphere flux making it to the top of the atmosphere. In \texttt{CLIMA}, the water vapor volume mixing ratio is fixed above the atmospheric cold-trap, and is directly dependent on the temperature of the tropopause. Case 1 converges on a tropopause temperature of 195\,K and adopts an upper atmospheric water vapor volume mixing ratio of $4.2\times10^{-6}$, more than double the amount of water vapor in the upper atmosphere that Case 3 converges on ($1.9\times10^{-6}$). The resulting water vapor absorption features in Case 3, are diminished compared to Case 1. However, the climate model adopts a similar amount of water vapor in the tropospheric region across all model states and these spectral models further indicate that tropospheric water vapor abundance is directly enshrouded by cloud properties.

\subsection{A Habitable Super-Earth Model}

\begin{figure}[ht]
    \centering
    \includegraphics{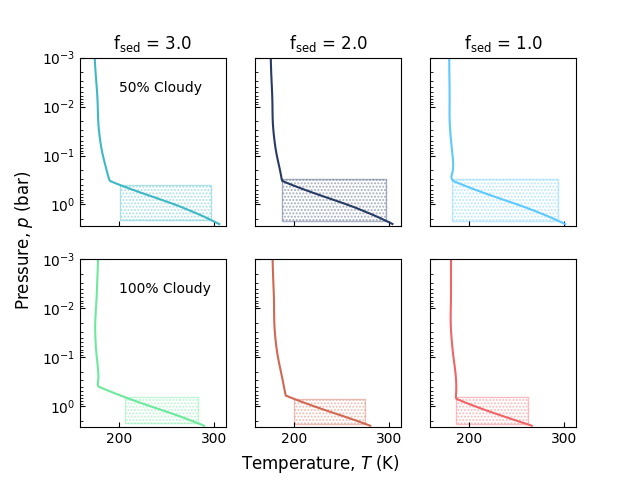}
    \caption{Super-Earth atmospheric thermal profiles for low (left column), med-range (middle column), and efficient (right column) values of the cloud sedimentation efficiency. Here, the top row represents 50\% cloudy climates and the bottom row showcases climates with 100\% clouds. Cloud vertical extent is indicated with a shaded box in the background.}
    \label{fig:Super_Earth_thermal_profile}
\end{figure}

\begin{figure}[ht]
    \centering
    \includegraphics{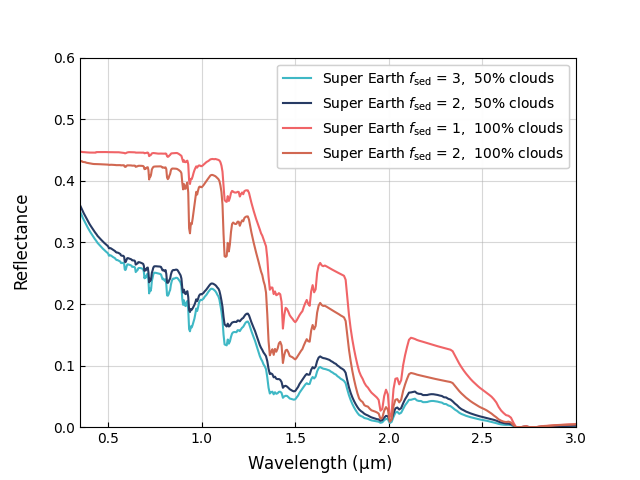}
    \caption{ \textbf{Top:} Reflectance spectra for the Super-Earth cases. Navy blue is is the Super-Earth climate with a mid-range sedimentaion efficiency light blue is a climate with the high sedimentaion efficiency. Here, Red represents a low range sedimentation efficiency of 1 with 100\% fractional cloudiness, and orange represents a mid-range sedimentation efficiency of 2. The Super-Earth models with 100\% clouds result in cloud decks that are comparably lower resulting in more water vapor absorption above the cloud layers and subsequently deeper water vapor absortion spectral features.}
    \label{fig:Best_Earth-like_Test_Cases_No_Labels_2}
\end{figure}

We use our validated self-consistent patchy cloud model to explore climate states of a super-Earth at 1\,au from a Sun-like host.  We adopt a total non-condensible species surface partial pressure of 2.5\,bar using Equation~(3) in \citet{kopparapuetal14} for a super-Earth with a radius of 1.5\,$R_{ \rm \bigoplus}$. Further, we adopt a super-Earth atmosphere comprised mostly of \ch{N2} (99\%), with  Earth-like mixing ratios of \ch{CO2} (3.6\,ppm) and \ch{Ar} (1\%), representative of an abiotic secondary super-Earth atmosphere \citep{Kite2020}. We explore sedimentation efficiencies of 1.0, 2.0, and 3.0 in the climate model resulting in Earth-like mid-cloud droplet radii of around 30, 60, and 120 \textmu m, respectively. Cloud coverages on habitable super-Earth planets are unknown, so we adopt basic scenarios of 50\% cloudy and 100\% cloudy. Finally, as earlier we fix the surface relative humidity at 100\%. The resulting super-Earth atmospheric thermal profiles are shown in Figure~\ref{fig:Super_Earth_thermal_profile}, and the resulting planetary reflectance spectra are shown in Figure~\ref{fig:Best_Earth-like_Test_Cases_No_Labels_2}. Our abiotic secondary atmosphere super-Earth models have no atmospheric oxygen/ozone source and the resulting thermal structures reflect this with relatively isothermal upper atmospheric structures.

\section{Discussion}
\label{sec:discussion}

The theory and results above detail a new 1D radiative-convective climate model for rocky worlds with patchy clouds. Comparisons between the model, Earth data, and more-complex tools help to validate our approach. Here, we discuss key details from model applications, especially with regards to cloud radiative effects, cloud structure, and super-Earth climates.


\subsection{Cloud Radiative Forcing}

\begin{figure}[ht]
    \centering
    \includegraphics{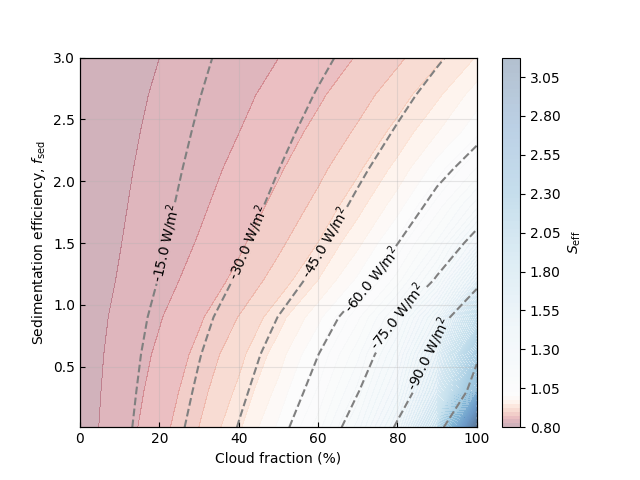}
    \caption{Cloud radiative forcing as a function of the sedimentation efficiency (vertical axis) and the cloud fraction (horizontal axis). The dashed gray contour lines correspond to cloud radiative forcing spaced by 15\,W\,m$^{-2}$, negative numbers correspond to the net cooling effect due to the presence of clouds. Warm (redder) colors correspond to climate states in which the overall atmosphere is undergoing a net warming effect. Conversely, cool colors (bluer) correspond to there the climate undergoes a net cooling effect. The white region centered between the -60\,W\,m$^{-2}$ and -75\,W\,m$^{-2}$ contour lines corresponds to where the net cooling effect of the clouds is balanced with the net warming effect due to water vapor absorption.}
    \label{fig:Cloud_Radiative_Forcing}
\end{figure}

\begin{figure}[ht]
    \centering
    \includegraphics{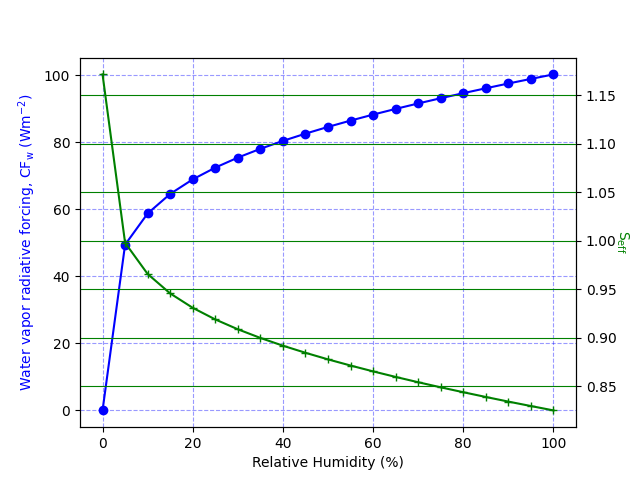}
    \caption{Water vapor radiative forcing (blue) and $S_{\rm eff}$ (green) as a function of tropospheric relative humidity. The radiative forcing of water vapor $CF_{\rm w}$ quickly approaches 50\,W\,m$^{-2}$ at a relative humidity value of 5\% throughout the troposphere. This corresponds to a $S_{\rm eff}$ value of 1.0 and represents a neutral radiative environment. Increasing the amount of water vapor results in higher radiative forcing values and an overall warming effect on the planetary surface.}
    \label{fig:water_vapor_radiative_forcing}
\end{figure}

Our gridded inverse climate study (Section~\ref{subsec:Exploring Cloud Parameters}) reveals strong climate state dependence on two main cloud model parameters\,---\,the amount of cloud cover and the overall sedimentation efficiency. These two parameters ultimately control the radiative environment in each atmospheric layer and are the main drivers of cloud behavior in our modeled planetary atmospheres. 

In order to isolate and explore the radiative effects of fractional cloudiness and sedimentation efficiency, we use this same set of gridded inverse models and employ the approach of \citep{Corti2009} to calculate the overall cloud radiative forcing in each individual climate. Here, 
\begin{equation}
    CF = CF_{\rm LW} + CF_{\rm SW} = (F_{\rm t}-F_{\rm t}^{\rm cld}) - (F_{\rm s} - F_{\rm s}^{\rm cld}) \, 
\end{equation}
where $CF$ is the overall cloud forcing value (typically with units of W\,m$^{-2}$), $CF_{\rm LW}$ is the long wave cloud radiative forcing, and $CF_{\rm SW}$ is the shortwave cloud radiative forcing, $F_{\rm t}$ corresponds to the longwave fluxes, and $F_{\rm s}$ corresponds to the shortwave fluxes. 

In general, the radiative forcing in both the long wave and the short wave can be calculated by selecting adjacent radiative transfer columns that share the same overall atmospheric structure (i.e. thermal profiles, species mixing ratios, atmospheric layer pressure) and subtracting the top-of-atmosphere clear-sky fluxes from the top-of-atmosphere cloudy fluxes.  In practice, this isolates any radiative effects that arise due to other atmospheric structures such as radiative interactions with other spectrally active gases like \ch{H2O}, \ch{CO2}, \ch{O3}, and \ch{O2}.

Figure~\ref{fig:Cloud_Radiative_Forcing} reveals that the overall cloud radiative forcing effect is negative. However, in order for clouds to have a net cooling effect on the planetary atmosphere, the cloud albedo effect must offset the net greenhouse effect. As can be interpreted from Figure~\ref{fig:Cloud_Radiative_Forcing}, the total cloud radiative forcing effect must grow to values larger than 65\,W\,m$^{-2}$  to offset the cumulative greenhouse warming. 

We quantify the overall planetary greenhouse effects of terrestrial \texttt{EddySed} clouds in Table~\ref{table:2} by measuring the global surface upwelling thermal flux and subtracting the top-of-atmosphere thermal flux. The resulting greenhouse fluxes are significantly higher than the greenhouse flux of the clearsky case (132\,W\,m$^{-2}$) due to the greenhouse contribution of water vapor \citep{pierrehumbert1995,marleyetal2010} in the clearsky sub-columns in addition to the cloud greenhouse flux. These high greenhouse fluxes result in climate states that are warm and, by definition, habitable \citep{kastingetal1993,kopparapuetal2013,kopparapuetal14} despite overall large planetary Bond albedos.


\begin{table}[ht]
\centering

\begin{tabular}{lccccc}
\hline
Climate Model                   & \multicolumn{1}{c}{Case 1} & \multicolumn{1}{c}{Case 2} & \multicolumn{1}{c}{Case 3} & \multicolumn{1}{c}{Case 4} & Tuned \texttt{CLIMA} \\ \hline
Cloud fraction, $f_{\rm c}$           & 100\%                          & 75\%                           & 60\%                          & 50\%                         &  n/a          \\ 
Sedimentation effiency, $f_{\rm sed}$                   & 3.0                            & 1.75                           & 1.0                           & 0.5                          & n/a          \\ 
Surface temperature, $T$ {(}K{)}     & 286                          & 284                          & 287                         & 288                        & 290        \\ 
Greenhouse flux {(}\,W\,m$^{-2}${)} & 250                          & 230                          & 234                        &226                         & 132        \\ 
\label{table:2}
\end{tabular}
\caption{Greenhouse flux comparison for Earth-like models.}
\end{table}

\subsection{Cloud Structure: Comparisons to Earth and Other Works}

Truly realistic Earth clouds cannot be completely represented through applications of a 1D convective cloud model like \texttt{EddySed}. Most importantly, 3D effects\,---\,such as advective mixing due to horizontal winds\,---\,cannot be realistically modeled in 1D. Further, precipitation on Earth can change the water vapor content throughout an atmospheric column \citep{manabe&wetherald1967,marleyetal2010} and, by comparison, our climate model assumes that the water vapor relative humidity profile stays uniform.

The \texttt{EddySed} framework places the cloud base at the point of saturation in a planetary atmosphere and the cloud structure is propagated upwards until layer saturation stops. In terrestrial planet atmospheres with water vapor as the main-condensible in their tropospheric regions, this scheme results in cloud structures throughout the entire convective region with vertical extents on the order of an atmospheric pressure scale height.  The large vertical extent of \texttt{EddySed} clouds are consistent with modeled water and ammonia clouds on Jupiter \citep{ackermanandmarley2001} where, in general, applications of \texttt{EddySed} to gas giant planets have resulted in cloud structures that span roughly a scale height within the much-deeper convective deep atmospheres of such worlds \citep{marleyetal2010}. It is likely that \texttt{EddySed} clouds are analogous to cumulonimbus cloud structures resulting from deep convective regions on the Earth \citep{Katoetal2020,Proud2021}, which comprise a smaller fraction of the overall cloud cover on Earth \citep{lacis2012}.

Previous 1D climate models for terrestrial planets with cloud treatments \citep{Kitzmann2010,Kitzmann2011,ZSOM2012} have taken a more heavily-parameterized approach to exploring the radiative effects of clouds. For example, \citet{Kitzmann2010} prescribes cloud physical properties such as number density, optical depth, and cloud top pressure with values from Earth measurements. These works have emphasized the importance of \textit{low-level} liquid water clouds and separate \textit{high-level} ice water clouds and are able to reproduce 1D flux profiles and globally averaged surface temperatures that are representative of true Earth.
This result requires overlap between the low-level and high-level clouds where, in some radiative transfer columns, the effects of both high-level clouds and low-level clouds are systematically incorporated. The radiative effects of the extended clouds produced by \texttt{EddySed} enable reasonably Earth-like climate states like those from more-tuned models. Overall the prescribed cloud structures in \citet{Kitzmann2010} \citep[and][]{Kitzmann2011,ZSOM2012} more realistically capture the cloud structure of the \textit{true} Earth and serve as a good comparison to the more-generalized \texttt{EddySed} cloud model adopted here.

\subsection{Habitable Super-Earths}

Super-Earths are likely to be among the most common type of world in our galaxy \citep{howardetal2012,fressinetal2013,Beanetal2021}. Yet in our own Solar System there are no analogs, leaving their composition and atmospheric structures to be inferred from observations \citep[e.g.,][]{kreidbergetal2014a} and a variety of planet models \citep{wordsworthetal2021}. Modeling results indicate that super-Earths can be habitable \citep{kopparapuetal14}, although habitability, here, takes the fairly narrow definition of surface thermal conditions appropriate for stable liquid water \citep{kastingetal1993,kopparapuetal2013,lincowskietal2018}.

We generate the first-ever 1D partially clouded super-Earth climate simulations that are paired to a high-resolution reflected-light spectral model. The resulting super-Earth atmospheric thermal structures suggest that the troposphere could be relatively Earth-like with surface temperatures around 300\,K at 1\,au from a Sun-like star and a 2.5\,bar N$_2$ atmosphere. These thermal conditions fall well within the globally-averaged surface temperature requirements for habitability \citep{kastingetal1993,kopparapuetal2013,ramirezetal2014,Godolt2016,wolf2017}. In our models, water vapor condensation is occurring throughout most of the troposphere (much like our baseline Earth-like model), cloud tops are truncated at the top of the convective zone, and the assumption of a non-oxygenated atmosphere results in a cold stratosphere whose temperature is near 180--190\,K for our suite of super-Earth models.

The super-Earth spectral models we generate show an expected trend in reflectivity as a function of overall fractional cloudiness, with a secondary trend corresponding to cloud droplet effective radii. 
Further study with this climate model should help illuminate linking spectral features from clouds to deep atmospheric environments. For instance, according to \citet{yangetal2013} planetary atmospheres with extremely low levels of cloud condensation nuclei may result in cloud droplet sizes that are double the value observed in Earth's atmosphere (corresponding to sedimentation efficiency values of 2.0--3.0 in our climate model). Thus, models with varying fractional cloudiness and sedimentation efficiency should be useful in rapidly exploring climate states, and associated observables, for a wide range of super-Earth exoplanets.

\subsection{Future Work}

The combination of \texttt{EddySed} with a widely-applicable rocky planet climate model, like \texttt{CLIMA}, will enable the study of a variety of wold types, the breadth of which would be a challenge to study with 3D models. For example, the condensate types inside \texttt{EddySed} are easily generalized, thereby enabling the study of cloud climates across a range of atmospheric conditions. An obvious area of study would be \ch{CO2} clouds in rocky planet atmospheres, where a number of studies have highlighted the importance of carbon dioxide condensation for impacting the outer edge of the habitable zone \citep{kastingetal1993,forget&pierrehumbert97,kopparapuetal2013,kopparapuetal14,kitzmann2017}

Another obvious application is to the inner edge of the Habitable Zone. \citet{yangetal2013} suggest that 3D cloud feedbacks for tidally locked planets on the inner edge of the Habitable Zone could move the inner edge limit to twice the flux level of the current cloud-free definition. Our cloud parameter study~\ref{subsec:Exploring Cloud Parameters} for a Sun-like host, indeed, suggests that there are some extreme climate states of Earth-like planets in which the inner limit of the Habitable Zone can extended to $2.2\times$ Earth's insolation. Our extreme case in which the Earth-like planet is 100\% covered in clouds coupled to a small sedimentation efficiency results in a $S_{\rm eff}$ value of 2.7 for an Earth-like (habitable) climate. This indicates a Habitable Zone inner edge that is well beyond the ranges in earlier works  \citep{kastingetal1993,selsisetal2007,Kitzmann2010,kopparapuetal2013,kopparapuetal14}. Similarly, it is apparent from the effective solar flux study that there are model cases where clouds can have a net warming effect on the surface, indicating that inner edge results will be varied. Finally, our study only considered Sun-like hosts and we expect there to be interesting interactions between cloud radiative properties and the host star spectrum that will lead to different cloudy inner edge trends with varying stellar host spectral type.

\section{Conclusion}
\label{sec:conclusion}

Clouds are near-ubiquitous in the Solar System and are of first-order importance for understanding exoplanet climates and observables \citep{hellingetal2019}. To enable the rapid study of rocky worlds with partially clouded atmospheres, we have developed a new 1D radiative-convective model for terrestrial planets that incorporates a parameterized treatment of patchy clouds. Our major findings are:

\begin{enumerate}
    \item Our patchy cloud radiative-convective model can produce Earth-like climate states where radiative fluxes are comparable to those determined by a full-physics, line-by-line radiative transfer model. 
    
    \item Sensitivity studies on the effect of \texttt{EddySed} cloud parameters on the overall climate of Earth-like atmospheres reveals a wide range of cloud states that can yield a habitable surface environment.

    \item While true Earth clouds cannot be completely represented in a 1D model, our coupled climate-cloud model balances complexity against computational efficiency, thereby enabling rapid parameter space explorations without the need for detailed cloud microphysics tools or 3D models.

    \item Our patchy cloud radiative-convective model reveals habitable super-Earth climates with tropospheric structures resembling that of our Earth and temperatures that imply that they are capable of hosting liquid water on their surfaces.
    
\end{enumerate}

\acknowledgements{TDR and JDW gratefully acknowledge support from NASA's Exoplanets Research Program (No.~80NSSC18K0349). TDR also acknowledges support from NASA's Exobiology Program (No.~80NSSC19K0473), the Nexus for Exoplanet System Science and NASA Astrobiology Institute Virtual Planetary Laboratory (No.~80NSSC18K0829), and the Cottrell Scholar Program administered by the Research Corporation for Science Advancement. R. K acknowledges support from the GSFC Sellers Exoplanet Environments Collaboration (SEEC), which is supported by NASA's Planetary Science Division's Research Program. J.L. acknowledges support from the National Science Foundation under Grant No AST2009343.}


%

\bibliographystyle{aasjournal}
\bibliography{biblist}

%

\end{document}